\documentclass[conference]{IEEEtran}
\IEEEoverridecommandlockouts
\usepackage{cite}
\usepackage{amsmath,amssymb,amsfonts}
\usepackage{algorithmic}
\usepackage{graphicx}
\usepackage{textcomp}
\usepackage{xcolor}
\def\BibTeX{{\rm B\kern-.05em{\sc i\kern-.025em b}\kern-.08em
    T\kern-.1667em\lower.7ex\hbox{E}\kern-.125emX}}
\begin{document}
 \makeatletter
\newcommand{\linebreakand}{%
\end{@IEEEauthorhalign}
\hfill\mbox{}\par
\mbox{}\hfill\begin{@IEEEauthorhalign}
}
\makeatother
\title{CORec-Cri: How collaborative and social technologies can help to contextualize crises? \\
}
\author{
%
		
		\IEEEauthorblockN{Ngoc Luyen Lê\IEEEauthorrefmark{1}\IEEEauthorrefmark{2}\\\textit{ngoc-luyen.le@hds.utc.fr}}\and
		\IEEEauthorblockN{Jinfeng Zhong\IEEEauthorrefmark{3} \\\textit{jinfeng.zhong@dauphine.eu} }
		\and
		\IEEEauthorblockN{Elsa Negre\IEEEauthorrefmark{3} \\\textit{elsa.negre@dauphine.fr} }
		\and
		\IEEEauthorblockN{Marie-Hélène Abel\IEEEauthorrefmark{1}\\\textit{marie-helene.abel@hds.utc.fr}}
		
		\linebreakand
		\IEEEauthorblockA{
			\IEEEauthorrefmark{1}Université de Technologie de Compiègne, CNRS, Heudiasyc (Heuristics and Diagnosis of Complex Systems),\\ CS 60319 - 60203 Compiègne Cedex, France.
			\\
			\IEEEauthorrefmark{2}Vivocaz, 8 B Rue de la Gare, 02200, Mercin-et-Vaux, France.
			\\
			\IEEEauthorrefmark{3}Paris-Dauphine University, PSL Research Universities, CNRS UMR 7243, LAMSADE, Paris, France.
		}
}
\maketitle

\begin{abstract}
Crisis situations can present complex and multifaceted challenges, often requiring the involvement of multiple organizations and stakeholders with varying areas of expertise, responsibilities, and resources. Acquiring accurate and timely information about impacted areas is crucial to effectively respond to these crises. In this paper, we investigate how collaborative and social technologies help to contextualize crises, including identifying impacted areas and real-time needs. To this end, we define CORec-Cri (Contextulized Ontology-based Recommender system for crisis management) based on existing work. Our motivation for this approach is two-fold: first, effective collaboration among stakeholders is essential for efficient and coordinated crisis response; second, social computing facilitates interaction, information flow, and collaboration among stakeholders. We detail the key components of our system design, highlighting its potential to support decision-making, resource allocation, and communication among stakeholders. Finally, we provide examples of how our system can be applied to contextualize crises to improve crisis management.
\end{abstract}


\section{Introduction}

In situations where traditional public resources like ambulances and helicopters are insufficient and not well-positioned to reach everyone in need during population evacuations, alternative evacuation resources must be explored. Citizen resources, including citizen-volunteer drivers and their personal vehicles, are more widely dispersed and accessible. Many citizen-volunteer may also be willing to assist with the evacuation using their own vehicles. For instance, a minivan owner with a capacity of 9 passengers could potentially evacuate an additional 8 people, significantly boosting the evacuation process's capacity. Similarly, a boat owner with a capacity of 6 passengers could help evacuate 5 people during a flood. Effective collaboration among different stakeholders is critical for a successful response to a crisis situation. In crisis management, collaboration involves working together with multiple stakeholders, including government agencies, non-governmental organizations, private companies, and individuals, to effectively respond to the crisis situation 
\cite{kapucu2011collaborative}. Key elements of collaboration in crisis management include communication, coordination, resource management, and decision-making. Effective communication is essential for successful collaboration in crisis management \cite{parker2020collaborative}. Stakeholders need to share information in a timely and accurate manner to make informed decisions and allocate resources effectively. Collaboration also involves coordinating efforts among stakeholders to avoid duplication of effort, make effective use of resources, and ensure that everyone is working towards common goals. Furthermore, collaboration requires stakeholders to work together to allocate and manage resources, including personnel, equipment, and supplies. Finally, stakeholders must work together to make informed decisions about response strategies and resource allocation. Furthermore, social computing has emerged as an essential tool for managing crises by facilitating information sharing, communication, and collaboration among stakeholders \cite{parameswaran2007social}. In the realm of crisis management, social computing has been applied to aid and relief efforts \cite{singh2019event,finch2016public,lim2016social}, crowd-sourcing \cite{alexander2014social,gao2011harnessing}, and managing information flow \cite{alexander2014social,nguyen2014critical,vihalemm2012citizens}.


We define an advanced crisis management system that extends the methodology introduced in \cite{le2023constraint} (ORec-Cri: Ontology-based Recommender system for Crisis management) by incorporating collaborative and social technologies to enhance communication and coordination between stakeholders. We term the modified system as CORec-Cri (Contextualized Ontology-based Recommender system for Crisis management). CORec-Cri recommends a list of citizen-volunteer drivers/vehicles to each impacted area, particularly when public resources are inadequate or inaccessible due to crisis situations. To further improve the system, we leverage social computing to better contextualize crises (e.g. identifying the location of impacted areas and real-time needs, gathering and sharing information related to crises) and facilitate collaboration among stakeholders. Collaborative technologies and social computing components are introduced to enable effective interactions among different stakeholders, supporting decision-making, resource allocation, and communication. We illustrate the practicality of our approach with examples of 
crisis situations where CORec-Cri can be applied. The key benefits of our approach include enhanced collaboration, informed decision-making, and improved crisis management outcomes.

The rest of this article is organized as follows: Section \ref{sec_02} introduces 
literature on collaborations and social computing in crisis management. Section \ref{sec_03} presents our main contributions: our approach to the construction of a framework leveraging the power of collaboration and  social computing for crisis management. In section \ref{sec_04}, we illustrate and discuss our work by a scenario of a crisis management
. Finally, we conclude and present the perspective.

\section{Related works}\label{sec_02}
In this paper, our aim is to contextualize crises through the utilization of collaborative and social technologies. Therefore, we will provide a brief overview of related works that leverage these technologies for effective crisis management.
\subsection{Collaborative Technologies in crisis management}
Collaboration plays an important role in crisis management, as it brings together various stakeholders and organizations to coordinate their efforts and respond to crises seamlessly. Effective collaboration in crisis management involves sharing information and resources, and making informed decisions that can impact the outcomes of evacuation activities. Collaborative technologies and social computing, combined with effective planning and training, can further facilitate collaboration, making it easier to share information and resources and prepare for crises  \cite{parker2020collaborative}. Therefore, the use of collaborative technologies can contribute to greater resilience in the face of crises and better preparedness for future unpredictable crises \cite{Chatham2016}.

Collaborative technologies based on computer-supported systems, networks and mobiles can provide an efficient means of enhancing the dissemination, retrieval, and analysis of information during crises.  For example, Monares et al. \cite{monares2009mobilemap} developed a mobile groupware application that enables firefighters to reduce their reliance on radio communication and exchange digital information during emergency response processes. Elmhadhbi et al. \cite{elmhadhbi2020promes} proposed an ontology-based messaging service that aims to resolve inconsistencies and enhance understanding among participants during emergency responses. This service ensures semantic translation of exchanged information, utilizing a shared understanding of concepts and terms among stakeholders. Chehade et al. \cite{chehade2020ontology} proposed the development of well-designed communication system interfaces that facilitate communication and interaction among various stakeholders, with the aim of enhancing situational awareness. Consequently, these proposed collaborative frameworks can improve the speed and accuracy of information sharing and analysis, which is crucial in crisis situations.

Collaborative technologies have several important aspects that can aid in organizing, deploying, and intervening during crisis management: (i) cooperative information gathering, (ii) team awareness, and (iii) risk and vulnerability assessment \cite{5990047}. Collaborative technologies first facilitate cooperative information gathering by enabling stakeholders to share and analyze critical data in real-time, which can help inform decision-making during evacuation efforts. Team awareness can then be enhanced by collaborative technologies that allow stakeholders to communicate and coordinate their efforts, ensuring that everyone is aware of the actions being taken. Finally, collaborative technologies can aid in risk and vulnerability assessment by providing stakeholders with the tools to identify and mitigate potential threats and vulnerabilities.
Messaging apps, social media platforms, and video conferencing tools are part of social computing that enable individuals to share information and communicate in real-time during crisis management. In the next section, we will examine related works on social computing and its applications in crisis management. 
\subsection{Social Computing in crisis management}
Social computing refers to the utilization of technology, such as Twitter and Facebook, to facilitate interaction and collaboration between users, transforming the way in which people communicate \cite{parameswaran2007social}. During times of disaster, whether natural or man-made, the use of social technologies rapidly increases. As Saroj and Pal \cite{saroj2020use} indicate, people use these technologies to communicate with family and friends in the affected area to inquire about their safety and security. Sharing or seeking information about essential needs such as food, shelter, transportation, and medical care is another important use of social technologies. With the help of social technologies, anyone can potentially serve as a source of essential information resources. Therefore, the use of such social technologies can be highly beneficial for decision-makers during times of crises \cite{havlik2016interaction, plotnick2015red, vieweg2010microblogging, alexander2014social}.

From existing literature, three ways can be identified to integrate social computing into crisis management. The first is the listening function \cite{alexander2014social}, which is a direct use of social computing since people impacted by crises can send out their opinions and emotions through social media. Emotional analysis during crises \cite{nguyen2014critical} has also been researched. Secondly, decision-makers can apply social computing to monitor the situation more effectively and respond efficiently to manage crises \cite{barr2011staying}. According to Vihalemm et al. \cite{vihalemm2012citizens}, social media helps citizens receive, understand, and emotionally cope with warning messages when crises arrive. The authors in \cite{singh2019event} used tweets to identify the location of crises, enabling decision-makers to better allocate rescue resources. Rescue or relief assistance is another critical use of social media tools by decision-makers, as medical needs can be identified more efficiently with the help of social media \cite{finch2016public,lim2016social}. Thirdly, crowd-sourcing and collaboration represent another important way social computing is applied to crisis management \cite{alexander2014social}. For instance, Sahana, along with its derivatives Eden, Vesuvius, and Mayon, are open-source disaster management systems that aim to harness the crowd-sourcing power of social media \cite{gao2011harnessing}. 

It can be observed that the use of social computing in these works focuses on private social media platforms such as Twitter, Facebook, etc. Although these platforms facilitate efficient information flow during crises, rumors and incorrect information may also spread and cause panic among the population, hindering effective crisis management \cite{panagiotopoulos2016social,han2019rumour}. In our work, we aim to design a social computing tool that is restricted to volunteers to avoid rumors and incorrect information and reduce the effort required to detect and mitigate the spread of such harmful information.
  
\section{Our Approach}\label{sec_03}
\subsection{Problem statement}
The authors in \cite{le2023constraint} presented a system (ORec-Cri) designed to aid decision-makers in recommending a list of citizen-volunteer drivers/vehicles for evacuation purposes. The system is composed of four layers: (i) the \emph{Interaction Layer}, where citizen-volunteer drivers respond to the requests of decision-makers; (ii) the \emph{Intelligent Layer}, which recommends a list of citizen-volunteer drivers/vehicles to decision-makers based on minimizing evacuation time; the (iii) \emph{Service Layer}, which computes the distance among points using \emph{OpenStreetMap}; (iv) the \emph{Data Layer}, which stores an ontology-supported knowledge base for the crisis management domain. However, there are two major issues with this system design. Firstly, it is not clear how the system collects critical information, such as the location of impacted areas, the number of people and disabled individuals in each area, and the priority level of each area. Secondly, the real-time situation in impacted areas may change, making it crucial that the system considers contextual information to generate accurate recommendations that adapt sto the real-time context of the impacted areas. To to address the issues of ORec-Cri \cite{le2023constraint}, we aim to tackle the following two problems: (i) how to collect detailed information related to impacted areas, such as location, number of affected individuals in each area, and the priority level of each area; and (ii) how to handle real-time changes in each impacted area, such as changes in accessibility. To solve these problems, we define a framework that harnesses the power of collaborative and social technologies.

\subsection{Collaboration framework for crisis management}
In the context of crisis management, we present a framework designed to facilitate effective collaboration and communication among various stakeholders  by leveraging collaborative and social technologies. This framework aims to improve the overall evacuation activities and population sheltering management based on the mobilization and deployment of citizen resources. The key aspects of our collaboration framework in crisis management are based on elements that encapsulate the common characteristics of collaboration as identified in the work of Sying et al. \cite{li2021context}. We have adapted these elements to the context of crisis management as follows: 
\subsubsection{Goals of collaborations} We aim to enhance the efficiency and effectiveness of evacuation efforts by fostering communication, coordination, and cooperation among stakeholders. These collaborations enable the exchange of accurate, timely, and relevant information, resource allocation, and decision-making, while collaborative and social technologies provide platforms and tools for real-time communication, data sharing, visualization, and tracking resources. Coordination and situational awareness
are also important goals in crisis management collaborations. With collaborative and social technologies and platforms, these objectives can be realized by using tools for planning, task assignment, progress monitoring, real-time data collection, analysis, knowledge sharing, capacity building, stakeholder engagement, and the documentation, analysis, and dissemination of knowledge. 
\subsubsection{Collaborators and Actions}
Collaborators in crisis management are essential to organize and participate in activities aimed at mitigating the impact of a crisis. Our research has identified several key collaborators, including decision-makers, citizen-volunteer drivers, and affected people. Regarding particular actions of the collaborators, decision-makers are essential to  coordinating and managing the overall evacuation operations. They are responsible for making critical decisions related to resource allocation for citizen-volunteer drivers/vehicles, task assignment to citizen-volunteer drivers, and overall strategy. This involves leveraging the collection and analysis of information, understanding the context of the crisis, and making informed decisions with the aid of technology. Citizen-volunteer drivers, on the other hand, are individuals who own or have access to vehicles and are willing to transport affected people to safe locations during a crisis. They may be called upon to help evacuate individuals from affected areas or transport essential supplies to those in need. Affected people are directly impacted by the crisis and may require assistance with evacuation, sheltering, or access to resources. This group includes individuals who may be injured, displaced, or otherwise affected by the crisis. Collaboration with affected people is crucial to ensure that their needs are identified and addressed in the response effort.

\subsubsection{Resources Applied} 
Effective crisis management requires the mobilization and coordination of different types of resources. In our scope, we categorize resources into three main types: material, human, and dematerialized resources. Material resources include physical tools, equipment, vehicles, and shelters necessary to respond to a crisis. This includes items such as citizen vehicles, search and rescue equipment, medical supplies, and emergency shelters. Human resources involve individuals involved in the response effort, including volunteers and affected individuals (who may need coordination with medical personnel or emergency responders). Dematerialized resources involve non-physical resources, such as data and information, that can be collected, analyzed, and shared to support decision-making and communication among stakeholders. The deployment of dematerialized resources can be the use of collaborative and social technologies to collect and analyze data and information in a timely manner, and share them with the relevant stakeholders to facilitate an efficient and coordinated response effort. By efficiently mobilizing and coordinating these resources, stakeholders can work together to respond to the crisis situation and minimize its impact on affected individuals and communities.
\subsubsection{Context of Collaborations}
Collaboration in crisis management requires consideration of various contextual factors that can impact the response effort. These factors may include access conditions, weather conditions, environmental conditions, and the time of day. Access conditions, such as the availability of transportation and communication infrastructure, can affect the ability of responders to reach affected people and the ability of affected people to access necessary resources. Weather conditions can impact the ability to respond or evacuate affected people, and environmental conditions such as the presence of hazardous materials can impact the safety of responders and those affected by the crisis. Additionally, the time of day can also be a factor, as evacuation efforts may differ depending on whether the crisis occurs during daylight or nighttime hours. Understanding the contextual factors can help decision-makers, affected people, and citizen-volunteer drivers collaborate more effectively by utilizing relevant contextual information to inform response strategies.

Effective collaboration in crisis management necessitates the cooperation of multiple stakeholders and the utilization of technological tools. Communication, coordination, and cooperation are prioritized throughout all stages of the response effort, making the construction of collaborative platforms crucial in facilitating this process and enhancing the overall response effort. 
We delve into the general architecture for developing computer-support systems for decision-makers during crises.


\begin{figure}[t]
  \centering
  \includegraphics[width=0.85\linewidth]{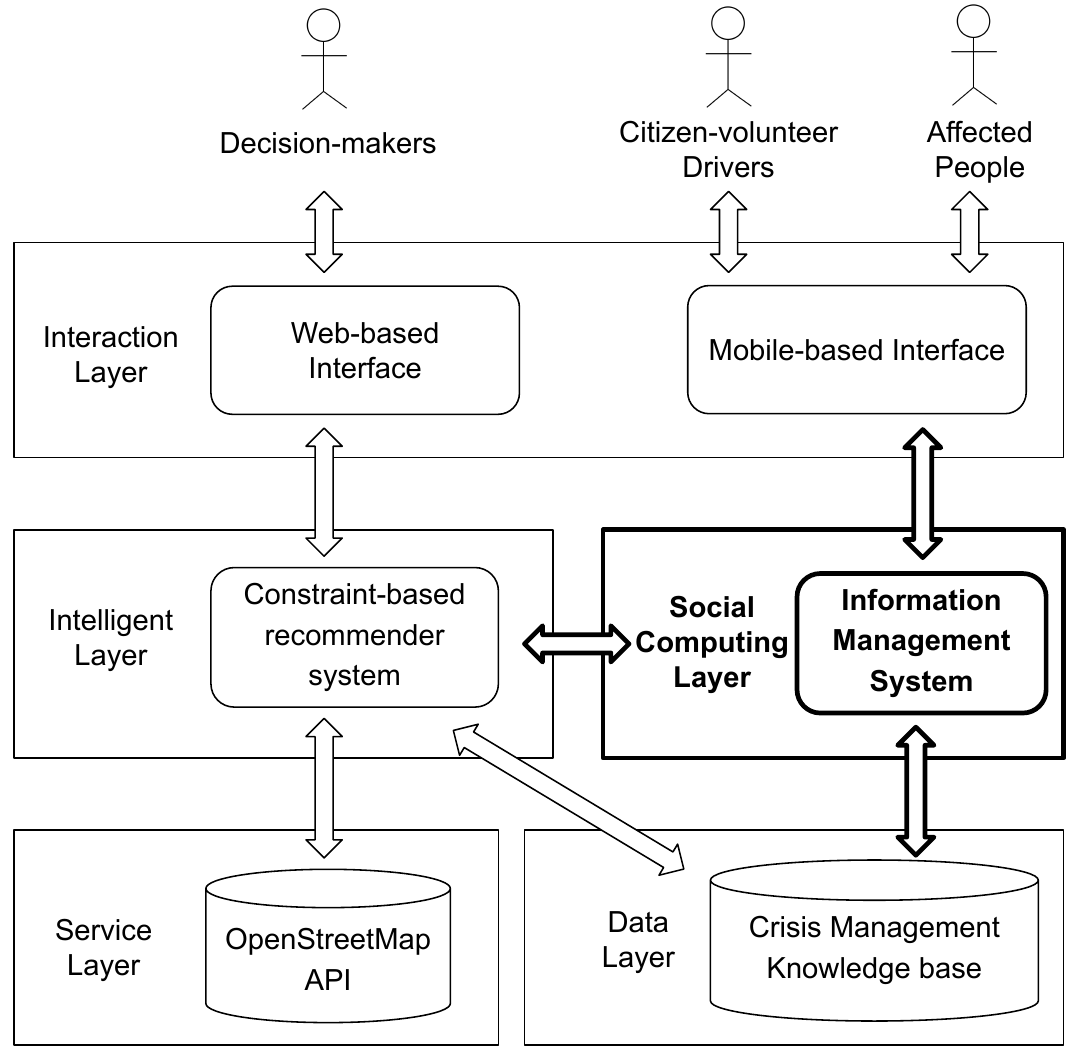}
  \vspace{-0.2cm}
  \caption{The architecture of CORec-Cri.}
   \label{fig1}
    \vspace{-0.8cm}
\end{figure}

The original architecture of ORec-Cri \cite{le2023constraint} is composed of four layers: \emph{Interaction Layer}, \emph{Intelligent Layer}, \emph{Service Layer} and \emph{Data Layer}. Firstly, \textit{Intelligent Layer} receives data from OpenStreetMap API and Crisis Management Knowledge base, and computes a list of citizen-volunteer driver/vehicle recommendations that satisfy the given constraints, which are displayed to decision-makers through the web-based interface. Then, \textit{Service Layer} computes the time for each distance between driver/vehicle pairs and rescue points and filters them accordingly. Next, \textit{Data Layer} contains an ontology-supported knowledge base for the crisis management domain, modeling and storing all necessary information and data \cite{le2023constraint}. By integrating of social computing into the current system, we define the construction of a novel architecture by adding a new layer called the \emph{Social Computing Layer}. This layer serves as an information management system during a crisis, and is responsible for collecting and processing data from different sources, such as affected people, citizen-volunteer drivers/vehicles, and decision-makers. The \emph{Social Computing Layer} will be able to integrate, exchange the data with other layers, making it easily accessible to decision-makers. As shown in Figure \ref{fig1}, the novel architecture for the platform includes the \emph{Social Computing Layer} as a central component for integrating and exchanging information during a crisis.

The integration of social computing into the proposed system can provide a powerful component for crisis management, allowing individuals, affected people, citizen-volunteer drivers/vehicles, and organizations to respond more effectively to crisis events by enabling real-time communication and collaboration, and providing valuable insights into how information is spreading and being received. Besides, integrating social computing can also enhance decision-making by providing direct information and feedback from other agents \cite{vivacqua2016collaboration}. This allows individuals, affected people, citizen-volunteer drivers/vehicles, and organizations to adapt their actions to the situations.

Establishing communication channels and platforms among various stakeholders in a crisis situation enables the real-time gathering of information about ongoing events. Decision-makers can capture and contextualize the situation from multiple perspectives, including those of the victims or affected populations, by continually updating information about the number of injuries or receiving reports from citizen-volunteer drivers/vehicles regarding any obstacles or accidents they encounter during rescue and evacuation operations. This collection of real-time information can significantly enhance the decision-making process and enable more efficient and effective participation in response efforts. Therefore, by utilizing collaborative and social technologies to contextualize crises, it is possible to achieve situational awareness, optimize resources, and make efficient decisions. The next section will explore the use of social computing in crisis management for context awareness.

\linespread{0.95}
\subsection{Contextualized crises by social computing}
Figure~\ref{fig1} illustrates the updated structure of our crisis management system with the addition of a \emph{Social Computing Layer} controlled by the \emph{Information management System}. In contrast to ORec-Cri proposed in \cite{le2023constraint}, CORec-Cri also allows impacted individuals (victims) to access the system via a \emph{Mobile-based Interface}. Through the integration of social computing, we aim to facilitate communication and collaboration among the various stakeholders involved in crisis management. This is crucial to contextualize crises and ensure better crisis management. In the following sections, we will provide a detailed overview of how the \emph{Social Computing Layer} helps to contextualize crises before, during, and after a crisis occurs: (i) \textbf{Before crisis:} citizen-volunteer drivers who are willing to help can register via the \emph{Mobile-based Interfaced}. Decision-makers may judge whether the registers are qualified to conduct rescue according to the information provided (e.g., rescue experience, detailed information related to their vehicles, etc.). For more details, please refer to \cite{le2023constraint}. This ensures that the citizen-volunteer drivers (with their vehicles) registered meet up with the constraints predefined. Plain users can  also register via the \emph{Mobile-based Interfaced} so that they can ask for help or release information about the impacted area when crises happen. (ii) \textbf{During crises:} In the event of an emergency, such as a flood, obtaining firsthand information about the impacted area quickly is crucial. Individuals impacted in these areas can assist decision-makers in identifying the location, the number of impacted individuals, and their actual needs (e.g., medical requirements, food, and water), which can be done through the \emph{Mobile-based Interface}. This information helps decision-makers recognize the priority of each impacted area and the real-time needs in the impacted areas. Once the needs have been identified, decision-makers can issue requests for assistance, and available citizen-volunteer drivers (with their vehicles) can respond through the \emph{Mobile-based Interface}. It is important to note that ORec-Cri utilizes the information provided by impacted individuals to recommend a list of citizen-volunteer drivers (with their vehicles) to the impacted areas. Additionally, social computing can be used to update the real-time situation in the impacted area, allowing individuals in the impacted area and citizen-volunteer drivers to assist decision-makers in better allocating various resources. (iii) \textbf{After crises:} The \emph{Social Computing Layer} not only facilitates communication and collaboration between the various stakeholders involved in crisis management but also enables decision-makers to release a synthesis report about the rescue process. This report can help citizen-volunteer drivers and impacted people to better understand how decisions were made and how resources were allocated. Additionally, citizen-volunteer drivers and impacted people can provide feedback, comments, and suggestions on the rescue process via the \emph{Mobile-based interface}. This feedback can help decision-makers to improve their crisis management strategies and processes. That being said, we can significantly improve real-time communication and collaboration between important factors in a crisis, such as victims/affected people, volunteers/drivers, and decision-makers. With the structure depicted in Figure~\ref{fig1}, the following three characteristics \cite{king2009brief} that capture the essence of social computing are assured: (i) \emph{Community:} In CORec-Cri, there are mainly three groups of people: decision-makers, affected people, and citizen-volunteer drivers/vehicles, which can be the collective source of wisdom \cite{king2009brief}. (ii) \emph{Connectivity:} In CORec-Cri, affected people, citizen-volunteer drivers/vehicles, and decision-makers are connected by the mobile-based interface. Affected people can inform decision-makers of real-time information about the affected place; citizen-volunteer drivers/vehicles can inform decision-makers of their availability; at the same time, decision-makers can inform affected people of the process of rescue. These interactions take place in the \emph{Intercation Layer} and are supported by the \emph{Social Computing Layer}. (iii) \emph{Collaboration:} As depicted in Figure~\ref{fig1}, the whole system operates by the collaboration of different people. Decision-makers help to allocate the citizen-volunteer drivers/vehicles;  citizen-volunteer drivers/vehicles help to rescue affected people; affected people keep decision-makers informed of the real-time information. At the same time, the self-organize efforts \cite{starbird2011voluntweeters} of  citizen-volunteer drivers/vehicles and affected people may also help to ease the shortage of resources. 

\linespread{0.95}
\section{Case Study and Discussions}\label{sec_04}
In this section, we describe a scenario where CORec-Cri can be implemented to manage a flood crisis situation in Compiègne city. The Municipal Council has taken measures to prepare for such a crisis by maintaining a list of 50 citizen-volunteer drivers/vehicles and three gymnasiums as shelters for vulnerable people. The information about these resources is entered into the system via a web interface.

When a flood crisis occurs, the person in charge at the Municipal Council initiates necessary evacuation activities. Decision-makers capture the number of people impacted, the priority level of each affected area, and the information about rescue points.  The mobilization of public resources has been deployed, but unfortunately, they do not suffice to cover all rescue points. Consequently, the person in charge has decided to reinforce the situation by utilizing citizen resources. Previously, the Municipal Council's agent utilized the mobile-based interface of CORec-Cri to gather information on available citizen-volunteer drivers and vehicle resources. Currently, decision-makers use a web-based interface to input information on the requirements of each rescue point, such as the number of affected individuals and the priority level of the rescue point, to obtain the best list of citizen-volunteer drivers and vehicle resources for each rescue point. Simultaneously, the person in charge can contextualize the crisis by employing the social computing module of CORec-Cri during the crisis, and they can monitor contextual information about the crisis, such as current weather and traffic conditions.

CORec-Cri creates a collaborative environment that enables decision-makers, citizen-volunteer drivers, and affected people to receive or share important information depending on their context. Specifically,  CORec-Cri utilizes a mobile-based interface to inform citizen-volunteer drivers of their assigned rescue point and the recommended trajectory. The system also facilitates real-time information sharing and contextual information, which allows decision-makers to monitor the situation of vehicles and the number of people present at each rescue point during the rescue process. By having updated, real-time information, decision-makers can make appropriate adaptations or interventions in response to unexpected changes in the situation. Citizen-volunteer drivers can also use the system to inform decision-makers of any changes in the rescue point, interruptions in traffic circulation, or inaccessibility to the rescue point. This information can assist decision-makers in making real-time adjustments to their plans

At the same time, the use of a social computing module can inform the public about the flood situation in Compiègne. Social platforms can capture specific situation emergences that may not be reported through traditional channels. In particular, incorporating the information management system and media platforms such as Twitter or Facebook, the Municipal Council can communicate with the public and provide updates on the situation, including the status of evacuation efforts, the availability of emergency resources, and other important information. Additionally, individuals impacted by the flood can use these platforms to request assistance or share information about their needs, allowing decision-makers to allocate resources more effectively. 


The use of collaborative and social technologies facilitates the real-time gathering of information about the situation and enables effective communication and coordination among stakeholders. CORec-Cri has the potential to assist decision-makers in managing a flood crisis situation efficiently and effectively by providing real-time information sharing, optimizing resources, and facilitating communication among decision-makers, citizen-volunteer drivers, and the public.
\section{Conclusion and Perspective}\label{sec_05}
In this paper, we have presented a collaboration framework for crisis management that utilizes collaborative and social technologies to enhance communication and coordination among various stakeholders. Through CORec-Cri, decision-makers, citizen-volunteer drivers, and affected people can leverage the benefits of social computing to gather, share, and analyze information about crises. This can help contextualize crises — before, during, and after a crisis occurs — and provide stakeholders with a better understanding of the situation, enabling them to share information in a timely and utilize resources effectively and efficiently. We have also illustrated the practicality of our approach with examples of how it can be applied in real-world crisis situations. 
Moving forward, there are several avenues for further research and development in the field of crisis management. First, there is a need for more robust and scalable systems that can handle large-scale crises with diverse requirements and constraints. CORec-Cri can be further enhanced by incorporating advanced technologies such as artificial intelligence, machine learning, and data analytics to improve decision-making, prediction, and resource allocation in real-time. Second, there is a need for a more comprehensive evaluation and validation of crisis management systems in real-world scenarios. Field trials and case studies can provide valuable insights into the effectiveness and efficiency of the system in different crisis situations and contexts. This can help in identifying potential limitations, areas for improvement, and opportunities for customization based on specific needs and requirements.

\section*{Acknowledgment}
This work was funded by the French Research Agency (ANR) and by the company Vivocaz under the project France Relance - preservation of R\&D employment (ANR-21-PRRD-0072-01).
\vspace{-0.2cm}
\bibliographystyle{ieeetr}
\bibliography{references}

\end{document}